\documentclass[conference]{IEEEtran}
\IEEEoverridecommandlockouts
\usepackage{preamble}
    
\begin{document}

\title{Securing the Management Plane in Intent-based Cellular Networks}

\author{\IEEEauthorblockN{Kashif Mehmood, Katina Kralevska, and Danilo Gligoroski} 
\IEEEauthorblockA{\textit{Department of Information Security and Communication Technology (IIK)} \\
\textit{NTNU\textemdash Norwegian University of Science and Technology, Trondheim, Norway}\\
Email: \{kashif.mehmood, katinak, danilog\}@ntnu.no}
}

\maketitle

\begin{abstract}
\Gls{ibn} is an emerging network management paradigm that allows automated closed-loop control and management of network devices and services. This paper addresses the security aspects in \gls{ibn} systems by securing the management plane in \Gls{ibn} systems. We propose a novel security framework based on WireGuard that augments the existing standards to secure intent communication between intent stakeholders. The framework guarantees isolation through WireGuard tunnels and provides inherent authentication and access control mechanisms to avoid intrusion in \gls{ibn} systems. Experimental results show the suitability and superiority of WireGuard compared to OpenVPN. 
\end{abstract}

\section{Introduction}
\Gls{ietf}, \gls{3gpp}, and \gls{tmf} have proposed different functional models \cite{tmf-intent-ig1253} for realizing an \gls{ibn} for a diverse set of network and service management applications. The increased reliance on automation and abstraction in \gls{ibn} introduces new security challenges. Ensuring the integrity and confidentiality of the network intents and the data traversing the network becomes critical. In this context, securing the management and data planes of \gls{ibn} systems is crucial to prevent unauthorized access, tampering, and other malicious activities that could compromise network operations.\par


The security in the management plane ensures that the decisions based on the intents are according to the needs of the intended consumers. Several works highlight the importance of implementing security functions in \gls{ibn} systems~\cite{kim_ibcs_2020, wang_intent-driven_2023, popova_security_2023}. However, there has not been any significant investigation regarding the importance and role of isolating the management plane of \gls{ibn}. \Gls{wg}~\cite{wg-Donenfeld2017}, a modern \gls{vpn} protocol, presents a compelling solution for enhancing security within \gls{ibn} environments. It provides simplistic, high-performance, and robust cryptographic assurances to address the security requirements of IBN management. 
\par

We propose a security framework based on the existing \gls{ibn} standards to secure the communication of intents between different intent stakeholders in a cellular network environment \cite{TS28.312_LTE, TR28.912_5G}. This paper presents a novel approach for enhancing security within the intent lifecycle by integrating authentication and access control between the \gls{ibn} stakeholders. The solution leverages \gls{wg} and a customized HTTPS-based key exchange mechanism to ensure secure, efficient, and flexible communication between \gls{csc}, \gls{csp}, \gls{nop}, \gls{visp}. The delegation between the intent stakeholders is managed by an \gls{ibnsc} that is also responsible for routing different intents between intent sources and providers. \gls{wg} is established among the intent stakeholders to secure the management plane, and experimental analysis shows \gls{wg}'s superiority over OpenVPN. 

The rest of the paper is organized as follows: Section II establishes the state-of-the-art about the security of \gls{ibn} systems. Section III describes the proposed security framework for \gls{ibn} with WG and Noise protocol. Section IV describes the experimental design and performance evaluation for the proposed WG-based \gls{ibn} system. Section V concludes the paper.


\section{Related Work}
\subsection{IBN Architecture}
\Gls{ietf} defines the basic architecture for implementing intent processing and translation capabilities in traditional networks \cite{IETF_IBN_NetMgmt}. Clemm \textit{et al.} \cite{RFC9315_IBN} present an abstract view of the intent lifecycle with mandatory processes such as intent definition, translation, and assurance. However, there is a lack of discussion on the management framework of intents in traditional networks and security requirements for different stakeholders. \Gls{tmf} proposes an \gls{imf} as a basic building block for implementing the different parts of the intent lifecycle \cite{tmf-intent-ig1253}. The processing and deployment of intents are based on the comprehension of available knowledge, decisions based on this knowledge, and the actions required to fulfill the objectives of the intent.\par

Similarly, \Gls{etsi} proposes an implementation framework for \gls{ibn} based on purely service-based architecture by treating each intent as a \gls{mns} \cite{TS28.312_LTE}. A \gls{mns} consumer generates the intent, whereas a \gls{mns} producer is responsible for fulfilling the intent objectives. For example, \gls{csc} generates an intent that must be fulfilled by the \gls{csp}, and \gls{csp} generates a more precise intent to be fulfilled by the \gls{nop}. The intents are categorized according to the scope as:
\begin{itemize}
    \item \textbf{Intent-CSC}: Intent for CSP with specifications for the required service, for example, "\textit{Enable remote industrial control of Area A}." 
    \item \textbf{Intent-CSP}: Intent for NOP with specifications for the networking service, for example, "\textit{Provide a network capable of supporting industrial control between Area A and rest of the network}."
    \item \textbf{Intent-NOP}: Intent for infrastructure or network equipment providers with information about specific network requirements, for example, "\textit{Provide a radio access network for Area A and core network capable of supporting ultra-reliable communications}."
\end{itemize}
A \gls{mns} consumer owns a generated intent, and a \gls{mns} producer is the handler of the intent \cite{10008673}. We utilize this terminology to develop a secure \gls{ibn} framework.

\subsection{IBN Security}
\textit{Kim et al.}~\cite{kim_ibcs_2020} proposed Security-as-a-Service for cloud-based service orchestration using intents. Security intents are translated and mapped to relevant virtualized security functions available in the cloud platform. Similarly, \textit{Wang et al.}~\cite{wang_intent-driven_2023} utilize network slicing to isolate different security intents and map them onto slice-specific capabilities. The network slicing process handles the intents' isolation and security along with the intents' lifecycle.\par
The problem of secure design for \gls{ibn} systems is handled by \textit{Ooi et al.}~\cite{ooi_intent-driven_2023}. The integration of security considerations within the intent lifecycle is investigated in different phases, from expression to deployment. Moreover, \textit{Chowdhary et al.} \cite{chowdhary_intent-driven_2022} propose a framework for defining a unified language model for security intent specification. However, the problem of securing the intent management plane is mainly unaddressed, leading to a considerable gap in understanding the security challenges.

\section{Proposed IBN Security Solution}
This section establishes the security challenges in IBN systems and presents the proposed model for deploying a \gls{wg}-based security solution for \gls{ibn}.\par
\begin{figure}[ht!]
  \centering
\includegraphics[width=\columnwidth]{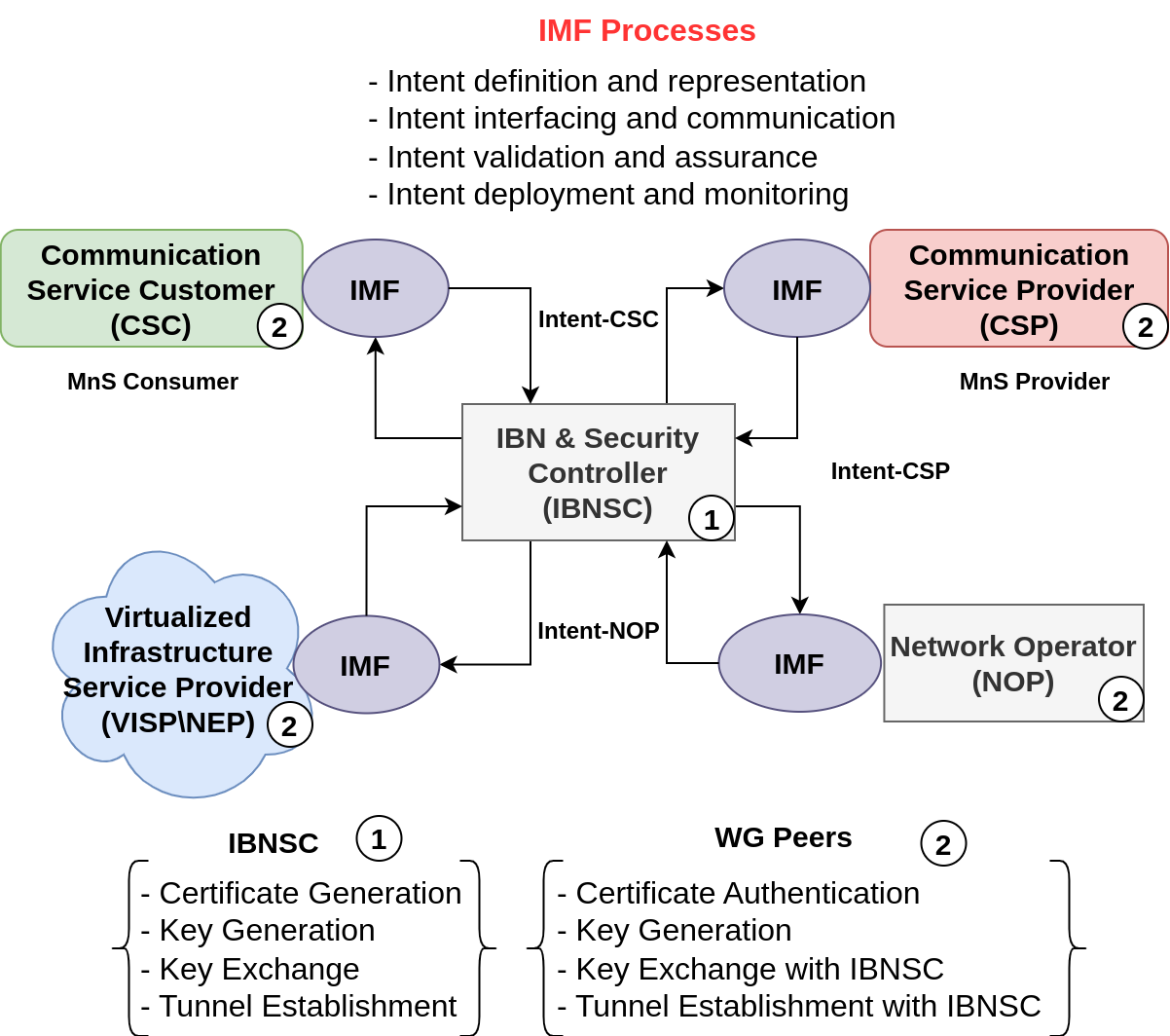}
  \caption{Security functions in the IBN management plane.}
  \label{fig:ibn-sec-roles}
\end{figure}
\subsection{Security Challenges in the Management Plane of IBN}
\Gls{imf} enables the communication and processing of various intents as a management service between \gls{ibn} stakeholders. Required resources are subsequently allocated by the \gls{csp}, \gls{nop}, and \gls{visp} to fulfill the objectives of the intents \cite{MEHMOOD2023109477}. The process is prone to intrusion and vulnerability due to the lack of an authorization and access control mechanism with the implementation of \gls{imf}.\par
Modern VPN solutions such as \gls{wg} \cite{wg-Donenfeld2017} can provide the required functionality in \gls{ibn} management and data plane. It supports various cryptographic methods such as ChaCha20, Ploy1305, Curve25519, and Blake2s for encryption, authentication, key exchange, and hashing \footnote{From a standardization point of view, we note that NIST has standardized Curve25519 in 2020 \cite{NIST-EC-2023}. The cryptographic community has scrutinized the other three algorithms; they are used in different IETF protocols but not standardized.}.
\subsection{Hybrid Management Model for IBN}
The proposed \gls{ibn} system utilizes the intent stakeholders defined by \gls{etsi}~\cite{TS28.312_LTE} suitable for mobile communication cellular networks. Moreover, the intent lifecycle definition is inspired by the \gls{tmf}'s intent specification~\cite{tmf-intent-ig1253}. A simple management model for intent stakeholders is shown in \figurename~\ref{fig:ibn-sec-roles} with the proposed hybrid \gls{ibnsc}.
\Gls{ibnsc} enables communication of intents between intent stakeholders. The intent stakeholders are embedded with \gls{imf} to support \gls{ibn} operation. \Gls{imf} provides the necessary lifecycle processes to translate, deploy, and assure the performance of the intents between relevant stakeholders \cite{tmf-intent-ig1253}.\par 
However, we focus on creating a secure management plane to implement and communicate intents. The security framework is centrally managed by the \gls{ibnsc}, whereas the \gls{imf} functionality is offloaded to individual intent stakeholders for context-centric understanding and deployment of intents. 
\subsection{Security Model for IBN}
The proposed security model for \gls{ibn} systems consists of a centralized \gls{ibnsc} responsible for authentication, access control, and secure propagation of intents amongst intent stakeholders as shown in \figurename~\ref{fig:ibn-sec-roles}. We propose an \gls{ibn} management plane implementing a \gls{wg} handshake mechanism based on the Noise protocol~\cite{noiseprotocol}. The reason for choosing \gls{wg} is embedded in its simplistic implementation and inherently superior cryptographic and handshake mechanism to competitors such as OpenVPN and IPSec~\cite{wg-opvpn-comp}.\par
 \begin{figure}[ht!]
  \centering
  \includegraphics[width=0.5\textwidth]{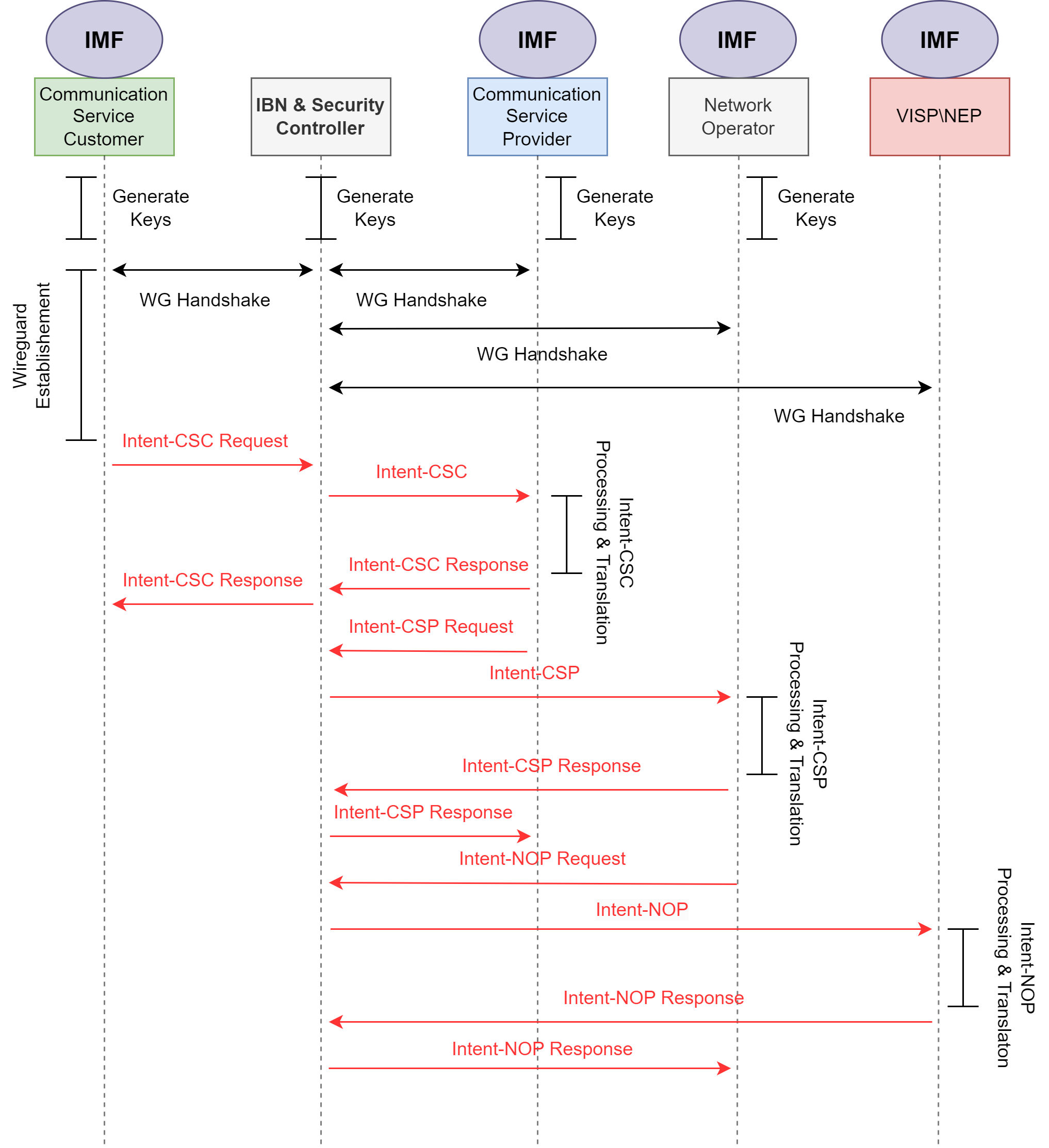}
  \caption{Proposed \gls{wg}-based IBN security protocol.}
  \label{fig:ibn-sec-wg}
\end{figure}
\subsubsection{Role of IBNSC}
\Gls{ibnsc} manages the intent interactions between stakeholders in an \gls{ibn} system. This includes routing intents from consumers to producers modeled as individual management services within the management plane. However, the global state of the intent and its assurance is available to the \gls{ibnsc} and utilized to measure the performance of the \gls{ibn} stakeholders and \gls{qos} delivery.\par
\Gls{ibnsc} also manages the security in \gls{ibn} by initiating the necessary procedures for authentication, access control, and cryptographic key establishment with different intent stakeholders. In the proposed architecture, \gls{ibnsc} acts as the authority that issues distinct authentication certificates to the intent stakeholders as \gls{wg} peers for key exchange.  We utilize specific security protocols for \textbf{Authentication} (Poly1305), \textbf{Key Exchange} (Curve25519), \textbf{Symmetric Encryption} (ChaCha20), and \textbf{Hashing} (BLAKE2s) to establish the \gls{wg} encrypted tunnels for intent communication between intent stakeholders. 
\subsection{Secure Intent Management Protocol}
We propose a key message exchange protocol for establishing intent communication between intent stakeholders using \gls{wg} encrypted tunnels with the Noise protocol. The proposed protocol consists of the following key steps as depicted in \figurename~\ref{fig:ibn-sec-wg}:
\begin{algorithm}
\caption{Key Exchange between CSC and IBNSC using Noise\textunderscore IK protocol}
\label{algo:noise_ik_hs}
\begin{algorithmic}[1]
\footnotesize
\setlength{\itemsep}{0.3em} 
\State \textbf{CSC} and \textbf{IBNSC} have static key pairs $(s_I, S_I)$ and $(s_R, S_R)$ \{$s_x$: private key, $S_x$: public key\}, where $x = \{I, R\}$
\State CSC knows $S_R$ in advance
\State Initialize: $ck = h = \text{BLAKE2s(prologue)}$
\State $h = \text{BLAKE2s}(h || \text{"Noise\_25519\_ChaChaPoly\_BLAKE2s"})$
\State CSC generates ephemeral key pair $(e_I, E_I)$
\State CSC computes:
\Statex $DH_1 = DH(e_I, S_R)$, $DH_2 = DH(s_I, S_R)$
\State CSC derives: $ck, temp\_k1 = \text{HKDF}(ck, DH_1 || DH_2, 2)$
\State CSC updates: $h = \text{BLAKE2s}(h || temp\_k1)$
\State CSC encrypts $S_I$ using $temp\_k1$ as key and $h$ as nonce
\State CSC $\xrightarrow{E_I, \text{encrypted}(S_I)}$ IBNSC
\State IBNSC generates ephemeral key pair $(e_R, E_R)$
\State IBNSC computes:
\Statex $DH_1, DH_2, DH_3 = DH(s_R, E_I)$, 
\Statex $DH_4 = DH(e_R, S_I)$, 
\Statex $DH_5 = DH(e_R, E_I)$
\State IBNSC derives: $ck, temp\_k2 = \text{HKDF}(ck, DH_3 || DH_4 || DH_5, 2)$
\State IBNSC updates: $h = \text{BLAKE2s}(h || temp\_k2)$
\State IBNSC $\xrightarrow{E_R}$ CSC
\State CSC computes $DH_3, DH_4, DH_5$
\State CSC derives: $ck, temp\_k2 = \text{HKDF}(ck, DH_3 || DH_4 || DH_5, 2)$
\State CSC updates: $h = \text{BLAKE2s}(h || temp\_k2)$
\State Both CSC and IBNSC derive final symmetric keys:
\Statex $k1, k2 = \text{HKDF}(ck, \text{zerolen}, 2)$
\State Initialize nonces: $n1 = 0$, $n2 = 0$
\end{algorithmic}
\label{algo:noise_hs}
\end{algorithm}
\subsubsection{Step 1 - Key Generation}
\gls{wg} uses asymmetric keys (static and ephemeral) and symmetric keys as implemented in Noise protocol \cite{wg-Donenfeld2017}. \gls{ibnsc} synchronizes public keys with its peers after the decentralized generation of asymmetric keys. This is established through a certificate authority at the \gls{ibnsc} responsible for authentication and secure establishment of asymmetric keys between \gls{ibnsc} and other intent stakeholders. The static keys $(s_x, S_x)$, where $x = {I, R}$ serve as the identity keys for authentication of different peers over a longer time. The ephemeral keys $(e_x, E_x)$, where $x = {I, R}$ are short-term session keys generated for each \gls{wg} session during the handshake initiation stage by the initiator and responder, respectively. The symmetric keys are used for data encryption (intents in IBN) transferred through the \gls{wg} encrypted tunnel.
\subsubsection{Step 2 - Authenticated Key Exchange}
The static public keys $S_I, S_R$ are exchanged using an \gls{https} server running at the \gls{ibnsc} after authentication is completed between \gls{ibnsc} and other intent stakeholders (CSC, CSP, NOP, and VISP). This key exchange authenticates the intended peers of the \gls{ibnsc} and establishes the \gls{wg} handshake procedure. 
The key exchange is performed using a modified Diffie-Hellman algorithm using Curve25519 cryptographic primitive.
\begin{figure*}[ht!]
    \centering
    \begin{subfigure}[b]{0.48\textwidth}
        \centering
        \includegraphics[width=\textwidth]{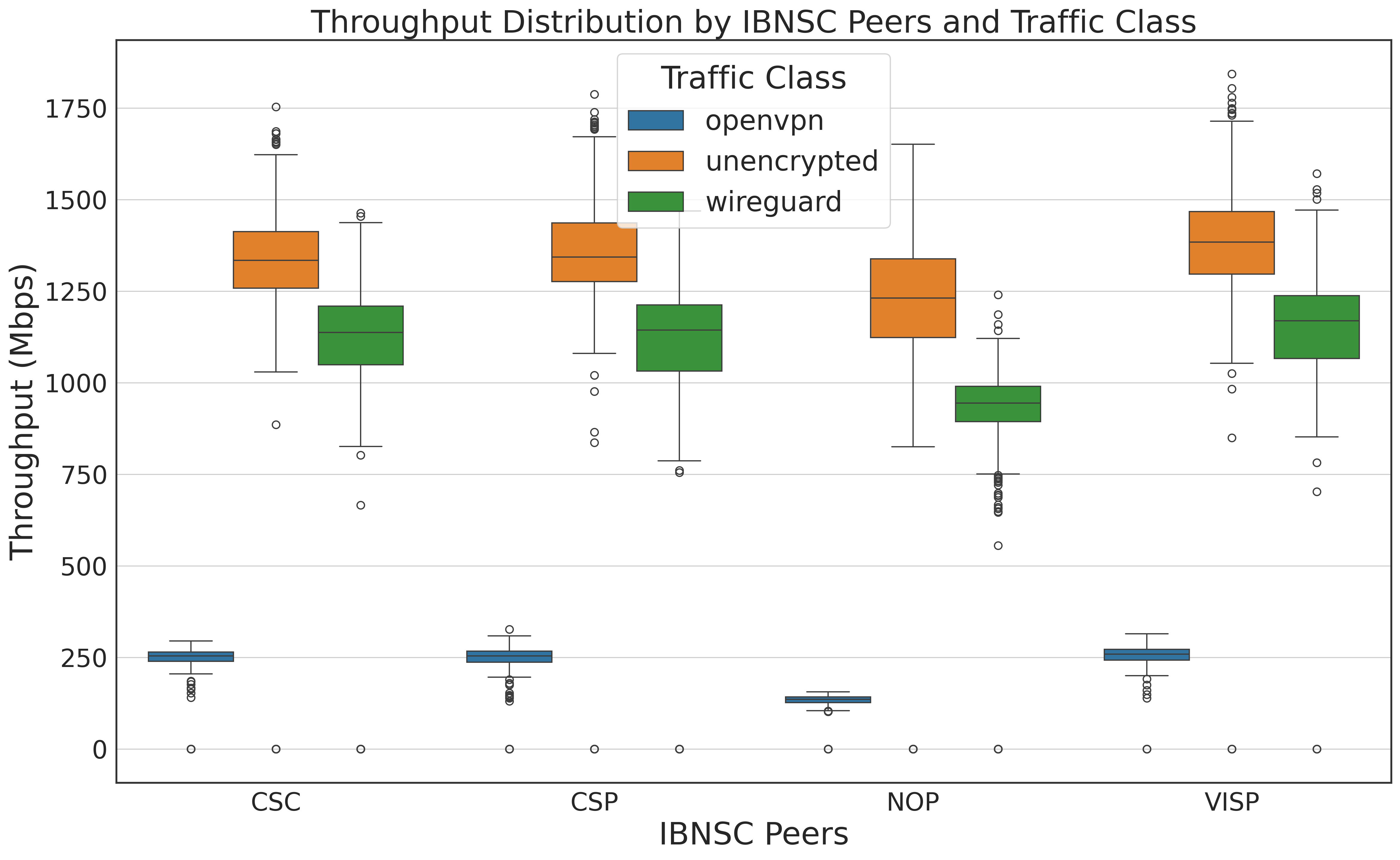}
        \caption{Throughput}
        \label{fig:tput-plot}
    \end{subfigure}
    \hfill
    \begin{subfigure}[b]{0.48\textwidth}
        \centering
        \includegraphics[width=\textwidth]{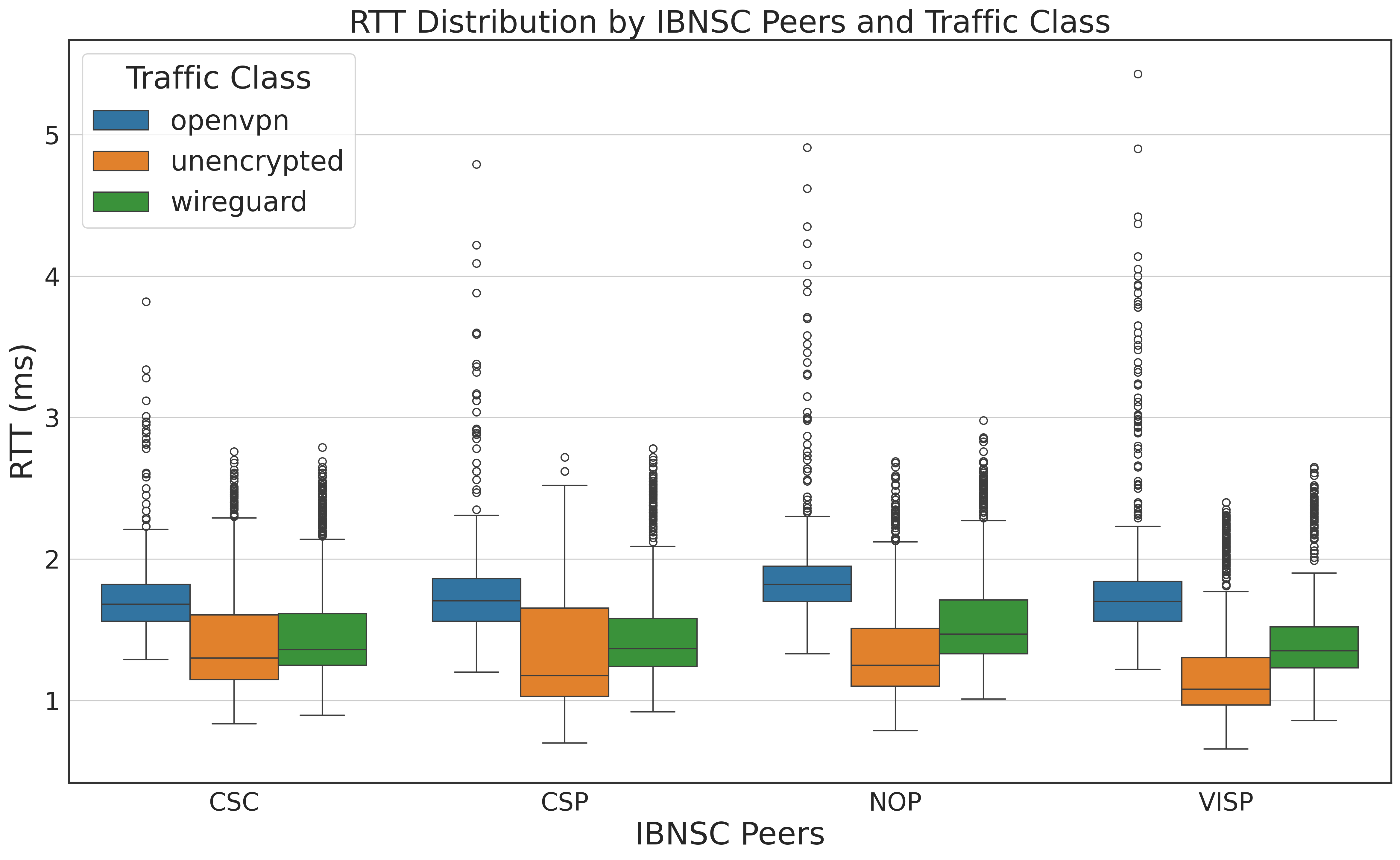}
        \caption{Round Trip Time}
        \label{fig:rtt-plot}
    \end{subfigure}
    \caption{E2E performance across IBNSC peers and traffic classes.}
    \label{fig:rtt-tput-perf}
    \vspace{-0.2cm}
\end{figure*}

\begin{figure*}[ht!]
    \centering
    \begin{subfigure}[b]{0.48\textwidth}
        \centering
        \includegraphics[width=\textwidth]{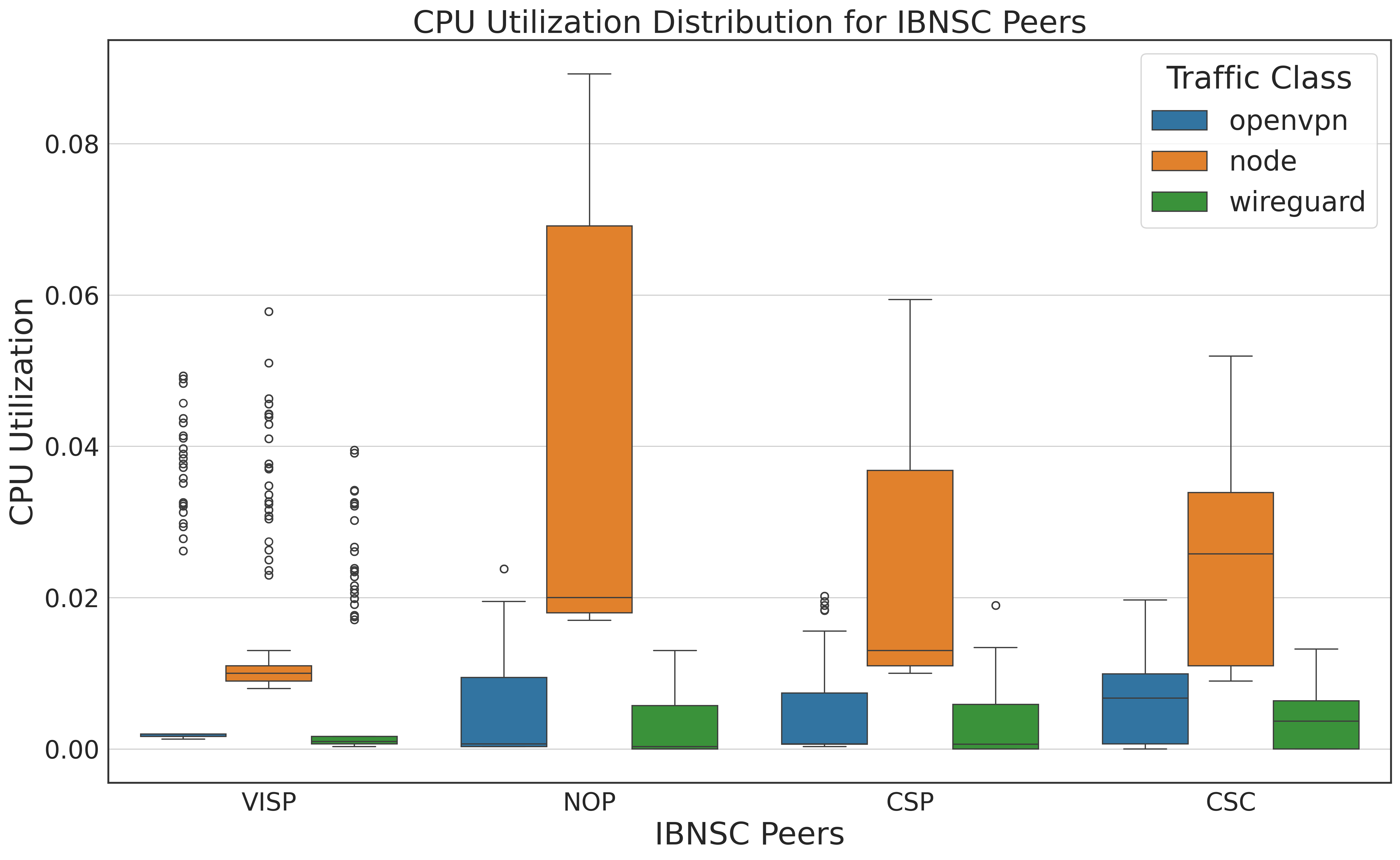}
        \caption{CPU}
        \label{fig:cpu-plot}
    \end{subfigure}
    \hfill
    \begin{subfigure}[b]{0.48\textwidth}
        \centering
        \includegraphics[width=\textwidth]{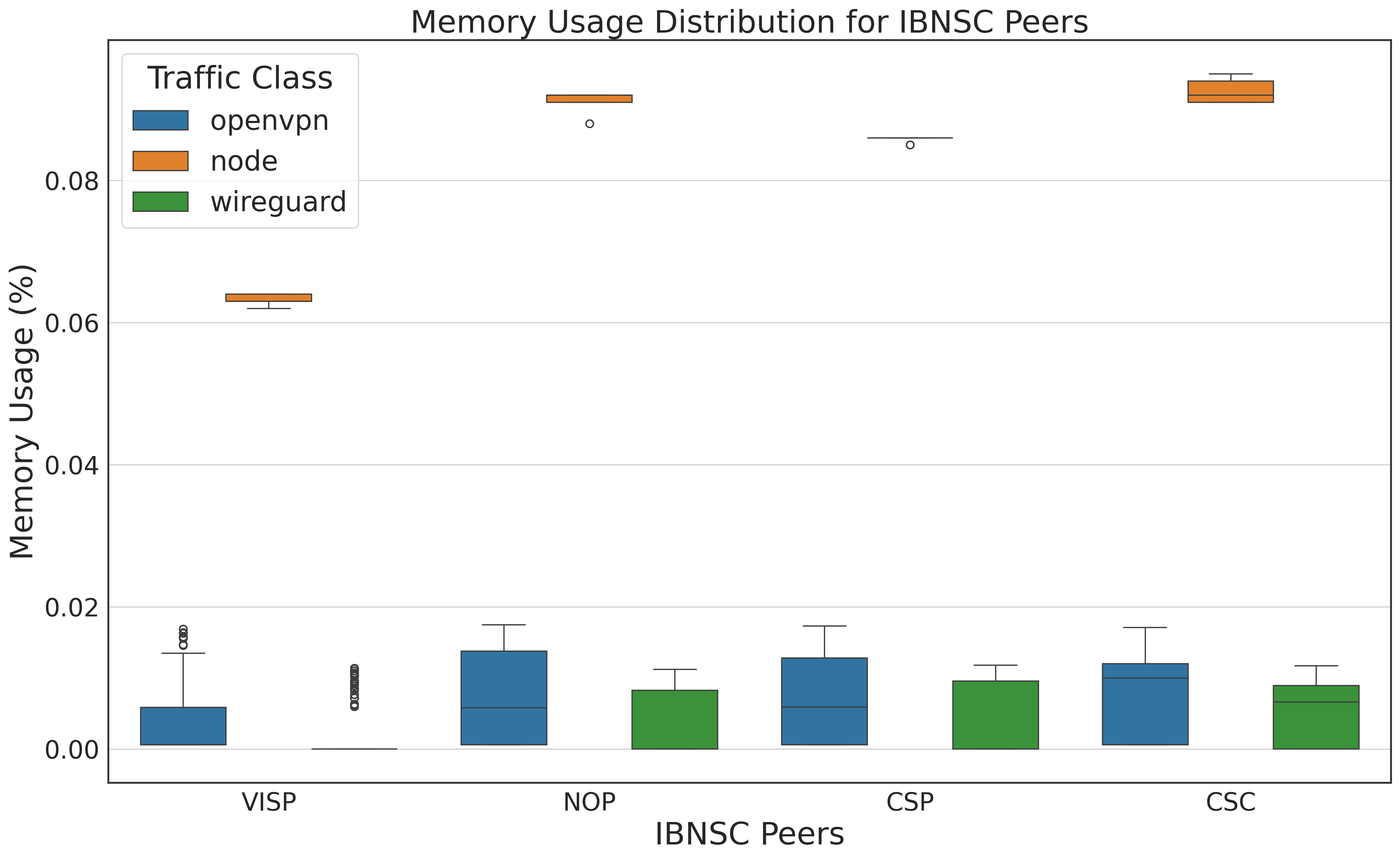}
        \caption{Memory}
        \label{fig:mem-plot}
    \end{subfigure}
    \caption{Node level metrics across IBNSC peers and traffic classes.}
    \vspace{-0.2cm}
    \label{fig:cpu-mem-perf}
\end{figure*}

\subsubsection{Step 3 - Handshake and Session Establishment}
Each intent stakeholder generates an asymmetric ephemeral key pair to initiate the handshake procedure with the \gls{ibnsc}. For example, \gls{csc} sends a handshake initiation message that includes its ephemeral and static public keys in the encrypted message using a shared secret generated from its known information. The responder \gls{ibnsc} sends the handshake response message with shared secrets using its ephemeral key pair and received keys. The handshake response message also includes the ephemeral public key of the \gls{ibnsc} as shown in Algorithm \ref{algo:noise_hs} \cite{noiseprotocol}.\par
The key exchange consists of Diffie-Hellman exchanges and HMAC-Key Distribution Function (HKDF) \cite{krawczyk2010hmac} coupled with the hashing function BLAKE2s \cite{aumasson2013blake2}. HKDF ensures strong encryption keys and uses the DH exchanges in the process.
\subsubsection{Step 4 - Intent Communication}
The \gls{wg} encrypted tunnel is active only between the authenticated peers and the \gls{ibnsc}. Intents can be communicated between different intent stakeholders via the \gls{ibnsc} in a secure channel using the symmetric secret keys generated with the information shared between \gls{wg} peers during the handshake procedure \cite{noiseprotocol}.\par


\section{Performance Evaluation}
\subsection{Experimental Setup}
The proposed experimental testbed consists of different intent stakeholders (\gls{csc}, \gls{csp}, \gls{nop}, \gls{visp}) modeled as individual \glspl{vm} connected via a shared network using OpenStack \gls{vim}. Each intent stakeholder is allocated 4 vCPUs, 16 GB of RAM, and 40GB of disk space. The connectivity between intent stakeholders and \gls{ibnsc} is established as follows: one virtual tunnel interface each is implemented in \gls{ibnsc} for connection using \gls{wg} and OpenVPN. One physical interface is implemented to record performance benchmarks for \textit{unencrypted} traffic without any established tunnels.
\subsection{Experimental Results}
The measurement probes and monitors are installed in the \gls{ibnsc}, and relevant metrics are recorded over several days using Prometheus. The experiments measure metrics for the resource (CPU and Memory) utilization of three individual interfaces: encrypted \gls{wg} interfaces, encrypted OpenVPN interface, and unencrypted interface established between intent stakeholders. The round-trip-time (RTT) and total throughput between \gls{ibnsc} and its peers are measured using iperf3.
\subsubsection{Throughput and RTT}
The results show that the \gls{wg} encrypted tunnel traffic provides higher throughput and lower RTTs than OpenVPN encrypted tunnel traffic between \gls{ibnsc} and its peers as shown in \figurename \ref{fig:rtt-tput-perf}. The interface with \textit{unencrypted} traffic achieves the highest useful data throughput between different intent stakeholders. However, it is followed closely by the \gls{wg} encrypted tunnel statistics, outperforming the performance of OpenVPN encrypted tunnel interface. The exceptional performance of \gls{wg} is attributed to several factors, including simplicity and lightweight design, cryptographic efficiency, stateless and optimized handshakes, and minimal overhead due to authentication and session establishment. These features provide the optimal security to communicate intents in \gls{ibn} suitable for low-latency and high-reliability application services. 
\subsubsection{Resource Utilization}
The resource utilization for the intent stakeholder node and VPN protocols is depicted in \figurename\ref{fig:cpu-mem-perf}. The \textit{node} level performance shows the overall resource utilization in individual intent stakeholders to serve as the baseline for the VPN services. The node level results also establish that the intent stakeholder VMs are not overloaded but operating under light load.\par
CSP has the highest observed load in the measurement duration with a peak CPU utilization of 9\%. \gls{wg} process in CSP has a maximum CPU utilization of around 1\%, approximately lower than the observed CPU utilization of 2\% for the OpenVPN process. A similar trend is observed, and \gls{wg} outperforms OpenVPN by requiring fewer CPU cycles for its associated processes.\par
Memory usage by OpenVPN and \gls{wg} follows an equivalent trend, with \gls{wg} requiring less memory resources than OpenVPN. The observed memory utilization for \gls{wg} remains below 0.01\% for all intent stakeholders.\par 

The functional aspects and security primitives used in \gls{wg} via Noise\textunderscore IK handshake pattern provide a simpler architecture with a 1-RTT handshake and low complexity, allowing minimal resource usage compared to other protocols \cite{wg-opvpn-comp}. Moreover, the low control overhead allows higher useful data throughput compared to OpenVPN encrypted tunnels between intent stakeholders. 


\section{Concusion}
This paper investigates possible security vulnerabilities in the \gls{ibn} management plane. \gls{wg} is poised as a suitable candidate to enhance the security, integrity, and reliability amongst intent stakeholders. The experimental evaluation provides insights into the suitability of the \gls{wg} protocol for \gls{ibn} systems. \gls{wg} achieved higher data rates and lower RTTs, corresponding to non-tunneled traffic. This efficiency is critical for ensuring low latency and high dependability services in IBN systems. The lightweight aspect of \gls{wg} was demonstrated by resource utilization data, which showed much lower CPU and memory usage across all intent stakeholders. These results verify our approach of employing \gls{wg} to secure the IBN management plane, providing a state-of-the-art solution that improves security without sacrificing performance. 

\bibliographystyle{ieeetr}
\typeout{}
\bibliography{refs}

\begin{thebibliography}{10}

\bibitem{tmf-intent-ig1253}
TMForum, ``{Intent} in {Autonomous} {Networks} v1.3.0,'' Tech. Rep. IG1253, TMForum, 2022.

\bibitem{kim_ibcs_2020}
J.~Kim, E.~Kim, J.~Yang, J.~Jeong, H.~Kim, S.~Hyun, H.~Yang, J.~Oh, Y.~Kim, S.~Hares, and L.~Dunbar, ``{IBCS}: {Intent}-{Based} {Cloud} {Services} for {Security} {Applications},'' {\em IEEE Communications Magazine}, vol.~58, pp.~45--51, Apr. 2020.

\bibitem{wang_intent-driven_2023}
K.~Wang, H.~Du, and L.~Su, ``Intent-{Driven} {Network} {Slicing} {Security} {Provision} and {Management},'' in {\em 2023 {IEEE} 23rd {International} {Conference} on {Communication} {Technology} ({ICCT})}, pp.~1318--1324, Oct. 2023.

\bibitem{popova_security_2023}
E.~Popova and D.~Lavrova, ``Security {Threats} to {Intent}-{Based} {Networks} in {Cyber}-{Physical} {Systems},'' in {\em Cyber-{Physical} {Systems} and {Control} {II}} (D.~G. Arseniev and N.~Aouf, eds.), (Cham), pp.~451--457, Springer International Publishing, 2023.

\bibitem{wg-Donenfeld2017}
J.~A. Donenfeld, ``Wireguard: Next generation kernel network tunnel,'' in {\em NDSS Symposium 2017}, 2017.

\bibitem{TS28.312_LTE}
{3GPP}, ``{5G; Management and orchestration; Intent driven management services for mobile networks},'' {Technical Specification} {TS 28.312 version 18.3.0 Release 18}, {3GPP}, 2023.

\bibitem{TR28.912_5G}
{3GPP}, ``{Study on enhanced intent driven management services for mobile networks},'' {Technical Report} {TR 28.912 version 18.0.1 Release 18}, {3GPP}, 2023.

\bibitem{IETF_IBN_NetMgmt}
{IETF}, ``{Intent-Based Network Management Automation in 5G Networks},'' 2023.

\bibitem{RFC9315_IBN}
{IETF}, ``{Intent-Based Networking - Concepts and Definitions},'' {RFC} 9315, {IETF}, 2022.

\bibitem{10008673}
M.~Xie, P.~H. Gomes, J.~Niemöller, and J.~P. Waldemar, ``{Intent-Driven Management for Multi-Vertical End-to-End Network Slicing Services},'' in {\em 2022 IEEE Globecom Workshops}, pp.~1285--1291, 2022.

\bibitem{ooi_intent-driven_2023}
S.~E. Ooi, R.~Beuran, T.~Kuroda, T.~Kuwahara, R.~Hotchi, N.~Fujita, and Y.~Tan, ``Intent-{Driven} {Secure} {System} {Design}: {Methodology} and {Implementation},'' {\em Computers \& Security}, vol.~124, Jan. 2023.

\bibitem{chowdhary_intent-driven_2022}
A.~Chowdhary, A.~Sabur, N.~Vadnere, and D.~Huang, ``Intent-{Driven} {Security} {Policy} {Management} for {Software}-{Defined} {Systems},'' {\em IEEE Transactions on Network and Service Management}, vol.~19, pp.~5208--5223, Dec. 2022.

\bibitem{MEHMOOD2023109477}
K.~Mehmood, K.~Kralevska, and D.~Palma, ``Intent-driven autonomous network and service management in future cellular networks: A structured literature review,'' {\em Computer Networks}, vol.~220, p.~109477, 2023.

\bibitem{NIST-EC-2023}
L.~Chen, D.~Moody, A.~Regenscheid, A.~Robinson, and K.~Randall, ``{Recommendations for Discrete Logarithm-based Cryptography: Elliptic Curve Domain Parameters},'' {SP 800-186}, {NIST}, 2023.

\bibitem{noiseprotocol}
T.~Perrin, ``Noise protocol framework.'' \url{http://www.noiseprotocol.org/}, 2023.

\bibitem{wg-opvpn-comp}
S.~Mackey, I.~Mihov, A.~Nosenko, F.~Vega, and Y.~Cheng, ``A performance comparison of wireguard and openvpn,'' in {\em Proceedings of the Tenth ACM Conference on Data and Application Security and Privacy}, CODASPY '20, p.~162–164, ACM, 2020.

\bibitem{krawczyk2010hmac}
H.~Krawczyk and P.~Eronen, ``{HMAC-based extract-and-expand key derivation function (HKDF)},'' tech. rep., {RFC 5869}, 2010.

\bibitem{aumasson2013blake2}
J.-P. Aumasson, S.~Neves, Z.~Wilcox-O'Hearn, and C.~Winnerlein, ``Blake2: Simpler, faster, secure hash function,'' ePrint 2013/322, IACR ePrint Archive, 2013.

\end{thebibliography}

\end{document}